\documentclass[]{iopart}
\usepackage{graphicx}

\newcommand{\sech}{\rm sech}
\newcommand{\be}{\begin{equation}}
\newcommand{\ee}{\end{equation}}
\newcommand{\bea}{\begin{eqnarray}}
\newcommand{\eea}{\end{eqnarray}}

\newcommand{\middlefig}{.5\textwidth}

\begin{document}

\title{Asymptotic Calculation of Discrete Nonlinear Wave Interactions}

\author{P.G. Kevrekidis$^1$, Avinash Khare$^2$, A. Saxena$^3$,
I. Bena$^4$ and A.R. Bishop$^3$}
\address{$^1$ Department of Mathematics and Statistics, University
of Massachusetts, Amherst, MA 01003-4515, USA}
\address{$^2$ Institute of Physics, 
Bhubaneswar, Orissa 751005, India}
\address{$^3$ Theoretical Division and Center for Nonlinear Studies, Los 
Alamos National Laboratory, Los Alamos, New Mexico 87545, USA}
\address{$^4$ Department of Theoretical Physics, University of Geneva, 
CH-1211 Geneva 4, Switzerland }

\begin{abstract}
We illustrate how to compute asymptotic
interactions between {\it discrete} solitary waves of dispersive
equations, using the approach proposed by Manton [Nucl. Phys. B
{\bf 150}, 397 (1979)]. We also discuss the complications arising
due to discreteness and showcase the application of the method 
in nonlinear Schr{\"o}dinger, as well as in Klein-Gordon lattices,
finding excellent agreement with direct numerical computations.
\end{abstract}

\maketitle

\section{Introduction}

Solitary wave interactions and collisions are among the 
signature characteristics of such nonlinear coherent structures
and have been extensively studied in both integrable and
non-integrable dispersive nonlinear 
models~\cite{mkdv1,AS,IR,Dodd,Scott,Rem}. The relevant physical contexts  
where such phenomena may arise range from the more traditional
areas of fluid mechanics and plasma physics (discussed in
the above references) or high-energy physics~\cite{belova}
to nonlinear optics~\cite{hasegawa,kivshar}, and the recently
emerging applications of Bose-Einstein condensates~\cite{stringari,nature}.

On the other hand, in the past decade there has been an explosive
parallel evolution of the theory and applications of {\it discrete}
nonlinear waves, namely discrete solitons and discrete breathers,
see, e.g.~\cite{reviews} for a number of review papers on the topic.
In this case, the applications refer to nonlinear waveguide
arrays~\cite{christonature} and photorefractive crystals~\cite{photo},
to Bose-Einstein condensates in deep optical lattices~\cite{BEC},
to the local denaturation of the DNA double strand~\cite{kim},
and to micro-mechanical cantilever arrays~\cite{sievers} among many others.

Our aim in the present work is to illustrate a technique for computing
solitary wave interactions in discrete systems. We should mention 
here that for continuum systems, there exists a large variety of
methods for computing such interactions. These range from 
the perturbation theoretical works of~\cite{karp}, to the variational
methods of~\cite{boris,todd,ricardo}, the Fredholm-alternative based
technique of~\cite{EMS} or the more rigorous calculations of~\cite{sandstede}
based on Lin's method. In a recent publication~\cite{avadh}, 
we took an alternative
route to these methods, by implementing an asymptotic calculation
using the approach proposed by Manton~\cite{manton} (which, in turn,
was generalizing the earlier work of ~\cite{perring}). Performing
such calculations in discrete systems is, however, considerably less
straightforward. One of the complications is that, aside from 
the tail-tail interaction of the waves, each wave may also reside  
on a periodic substrate potential, usually termed the Peierls-Nabarro
barrier~\cite{reviews}. As a result, except for the (typically
exponential in the wave separation) tail-induced potential,
a local potential may arise due to discreteness, resulting in a
washboard potential structure which can be captured by variational
techniques~\cite{todd,ricardo}.

In what follows, we show how to compute discrete solitary wave
interactions by means of Manton's method. We bypass the issue of the 
Peierls-Nabarro barrier by working with problems whose static 
solutions do {\it not} encounter such a barrier (so that they can be 
placed anywhere on the lattice).  We demonstrate this approach both 
for breathers in nonlinear Schr{\"o}dinger
lattices  -- namely in its famous Ablowitz-Ladik (ALNLS) variant~\cite{AL} --
and for kinks in Klein-Gordon lattices  -- namely in a discrete version
of the $\phi^4$ model, proposed in~\cite{bender,kevrek,cooper}. While
the former model is integrable, integrability is {\it not} a key
ingredient in our calculation. Instead, the existence of a 
{\it discrete analog of momentum conservation law} 
(and also the absence 
of the Peierls-Nabarro barrier) are indispensable for being able
to carry through the Manton calculation as we show in the following
sections. We compare our analytical results with direct numerical
computations, finding excellent agreement between the two during the
time interval in which the waves interact without losing their 
individual character. Furthermore, our calculations also capture the 
correct continuum limit (known from the earlier works of
\cite{boris,manton}; see also \cite{avadh}), as the lattice
spacing $h \rightarrow 0$.  

Our presentation is organized as follows. In section 2, we 
illustrate the method for the breathing solitons of the Ablowitz-Ladik
model. In section 3, we compute the interaction potential for
kinks in a non-integrable, discrete $\phi^4$ model. Finally, in
section 4 we summarize our results and present our conclusions
and some interesting directions for future study.

\section{Soliton Interactions in the Ablowitz-Ladik Model} 

Similarly to the corresponding continuum calculation (see, e.g.,~\cite{avadh} 
for such calculations in different models), the 
key to performing a calculation for the discrete case using Manton's method is the
existence of a momentum as a conserved quantity. In the ALNLS model, 
such a momentum operator is one of the integrals of motion (that 
is why integrability is desirable for performing such a calculation, 
even though, as we will show below, it is not a necessary condition), and has the 
form (see, e.g.,~ \cite{AS,cai} for details)
\begin{eqnarray}
P=i \sum_{n=-\infty}^{\infty} \left(\psi_n \psi_{n+1}^{\star} -
\psi_n^{\star} \psi_{n+1} \right),
\label{dmeq1}
\end{eqnarray}
where $\psi$ is the complex field, and the star is used
to denote complex conjugation.
We will focus here on the ALNLS equation in the form \cite{cai}
\begin{eqnarray}
i \dot{\psi}_n+(\psi_{n+1}+\psi_{n-1}) (1+ |\psi_n|^2) = 0 ,
\label{dmeq2}
\end{eqnarray}
with the explicit stationary soliton solution
\begin{eqnarray}
\psi_n=\sinh(\beta) {\rm sech}\left(\beta (n-s)\right) \exp(i \sigma) , 
\label{dmeq3}
\end{eqnarray}
where $\dot{\sigma}=2 \cosh(\beta)$. Here the overdot denotes time 
derivative, $\beta$ is the inverse width of the pulse soliton and $s$ 
is the arbitrary position of the center of the pulse.  
We commence by examining in-phase solitons, but we will also relax
this constraint later. Note that Eq.~(\ref{dmeq2})
has the lattice spacing $h$ scaled out, but it is straightforward
to incorporate it and we will see how to relate it to the relevant
continuum limit of $h \rightarrow 0$ for our calculations.

In the spirit of \cite{manton}, we now consider two solitons,
one centered at $0$ and one centered at $s \gg 0$, i.e., two 
{\em widely-separated} solitons. We compute 
$d P/dt$ by performing the summation over $n$ not for the infinite lattice
(when the result would be zero due to the relevant conservation
law), but rather from $n=M$ to $n=N$, with $M \ll 0$, and
$0\ll N \ll s$. The idea behind this calculation is that, in fact,
the force in this interval is not going to be zero, but rather
would be {\it finite} due to the soliton-soliton interaction.
When summing on the infinite line, as a result of Newton's third
law, the action of the first soliton on the second and the 
equal and opposite reaction of the second on the first cancel
each other, thus resulting in zero net momentum gain. However, for a finite
interval encompassing only  one soliton, the amount of momentum
gain is finite, due to the fact that the one soliton experiences  
the pull (or push) of the other soliton at the boundary of the
interval where we perform the calculation. In precise mathematical
terms, we evaluate:
\begin{eqnarray}
\frac{dP}{dt} &=& -2
\sum_{n=M}^{N} \left(|\psi_{n+1}|^2-|\psi_n|^2\right)
\nonumber
\\
&+& \sum_{n=M}^{N}
\left(\psi_n \psi_{n+2}^{\star}+\psi_n^{\star}\psi_{n+2}\right)
\left(1+|\psi_{n+1}|^2\right) 
\nonumber
\\
&-&  
\sum_{n=M}^{N} 
\left(\psi_{n-1} \psi_{n+1}^{\star}+\psi_{n-1}^{\star}\psi_{n+1}\right)
\left(1+|\psi_{n}|^2\right) .
\label{dmeq4}
\end{eqnarray}
However, observing the telescopic nature of the sums in the right hand
side (RHS) of Eq. (\ref{dmeq4}), we infer that 
\begin{eqnarray}
\frac{dP}{dt} &=&
-2 \left(|\psi_{N+1}|^2-|\psi_{M}|^2\right)
\nonumber
\\
&+& 
\left(\psi_N \psi_{N+2}^{\star}+\psi_N^{\star}\psi_{N+2}\right)
\left(1+|\psi_{N+1}|^2\right) 
\nonumber
\\
&-&  
\left(\psi_{M-1} \psi_{M+1}^{\star}+\psi_{M-1}^{\star}\psi_{M+1}\right)
\left(1+|\psi_{M}|^2\right) .  
\label{dmeq5}
\end{eqnarray}
As usual in Manton's method, and based on intuitive physical arguments, 
the main contribution in this asymptotic calculation stems from the 
boundary between the two solitons. Hence, we drop the terms
with subscript $M$ and only consider the contributions with subscript
$N$ in what follows.

We then invoke the soliton ansatz
\begin{eqnarray}
\psi_n=\psi_n^{(1)}+ \psi_n^{(2)}
\label{dmeq6}
\end{eqnarray}
with $\psi_n^{(1)}=\sinh(\beta) {\rm sech(\beta n)} \exp(i \sigma)$
and $\psi_n^{(2)}=\sinh(\beta) {\rm sech(\beta (n-s))} \exp(i
\sigma)$
(i.e., two in-phase solitons).  Since $0 \ll N \ll s$, we can use the asymptotic form of the soliton 
tail at $n=N$, according to
\begin{eqnarray}
\psi_n^{(1)}=2 \sinh(\beta) \exp(-\beta N) \exp(i \sigma)\, , 
\label{dmeq7}
\\
\psi_n^{(2)}=2 \sinh(\beta) \exp(\beta (N-s)) \exp(i \sigma)\, . 
\label{dmeq8}
\end{eqnarray}
Substituting the ansatz of Eq. ~(\ref{dmeq6}) and the expressions in
Eqs.~(\ref{dmeq7})-(\ref{dmeq8}) into Eq.~(\ref{dmeq5}), 
and computing the terms arising
from the soliton-soliton interaction, we obtain that 
\begin{eqnarray}
\frac{dP}{dt} \approx 32 \sinh^4(\beta) \exp(-\beta s)\,.
\label{dmeq9}
\end{eqnarray}

We note a number of comments on this calculation.
\begin{itemize}
\item {\em Equation of motion for $s(t)$}: As  explained in~\cite{avadh},
in order to obtain the equation of motion for the inter-soliton separation
$s(t)$,
we can use  Newton's equation in the form
\begin{eqnarray}
M_s \ddot{s}=-2 \frac{dP}{dt}\,,
\label{dmeq10}
\end{eqnarray}
where $M_s$ is the mass of the soliton; the factor  ``$2$" comes
from the fact that there is an equal and opposite pull (or push)
on the second soliton, and hence their relative distance decreases
by twice the contribution of $dP/dt$ to each of them; and finally 
the unfamiliar ``--" sign originates from the fact that a positive
boundary contribution to $dP/dt$ decreases the soliton distance,
while the opposite is true for a negative $dP/dt$. In the present
case, the soliton mass is given by~ \cite{cai}
\begin{eqnarray}
M_s=\sum_{n=-\infty}^{\infty} {\rm ln} \left(1+|\psi_n|^2\right)=2 \beta
\label{dmeq11}
\end{eqnarray}  (once again, a direct benefit
of integrability being the immediate accessibility of the relevant
conservation law).
Then one can straightforwardly infer the equation for $s(t)$ as
\begin{eqnarray}
\ddot{s}=-\frac{32}{\beta} \sinh^4(\beta) \exp(-\beta s)\,,
\label{dmeq12}
\end{eqnarray}
while the relevant effective soliton interaction potential 
(for a unit mass particle) is 
\begin{eqnarray}
V(s)=-\frac{32}{\beta^2} \sinh^4(\beta) \exp(-\beta s)\,.
\label{dmeq13}
\end{eqnarray}
We have examined the validity of the pertinent calculation by studying 
numerically the
two-soliton collisions in the ALNLS model. Typical results
are illustrated in Fig. \ref{dmfig1} for $\beta=2$ and two solitons
initialized at a distance of $s(0)=10$ between them. The upper panel
of the figure shows the numerical (solid line) versus theoretical (dashed line)
prediction for the soliton separation as a function of time, $s(t)$,
revealing excellent agreement for all times until about $t \approx 657$, when
the separation becomes $s<3$ and the solitons can no longer be 
characterized as individual entities; hence, it is 
reasonable that our asymptotic calculations fail at that point. In
fact, it appears that the analytical calculation gives a very good
agreement with the numerics well past the point where we might have
anticipated such agreement (on the basis of our expansion
assumptions), and approximately even up to the collision point. 

\begin{figure}
\begin{center}
\begin{tabular}{c}
    \includegraphics[width=\middlefig]{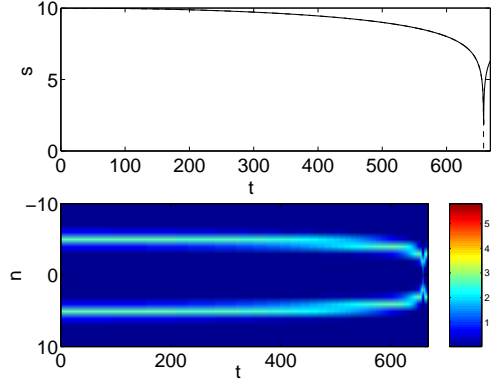}
\end{tabular}
\caption{The top panel shows the soliton-soliton separation as a function of
time, $s(t)$, for the numerical integration of the ALNLS equation with 
two superposed solitons (at $n=-5$ and $n=5$) as the initial
condition.
The solid line yields the numerical result (calculated by obtaining
independently the center of mass of the left and right pulse in
the configuration, and subtracting the former from the latter), while
the dashed line provides our asymptotic prediction of
Eq.~(\ref{dmeq12}). Note the  excellent agreement between the two,
essentially up to  the collision time, since
the two curves cannot  be distinguished, practically, up to that time. 
The bottom panel shows the
space-time contour plot of the ``local mass" $m_n=
{\rm ln}(1+|\psi_n|^2)$, clearly illustrating the evolution towards
the collision of the center of mass of the two solitons. }
\label{dmfig1}
\end{center}
\end{figure}

\item {\em Role of the soliton relative phase}:  In order to examine it,
it is simple to impart to one of the ALNLS solitons a free phase
$\exp(i \phi)$, with respect to the other soliton. In this case,
the results of Eqs.~(\ref{dmeq12})-(\ref{dmeq13}) are retrieved,
but with the RHS expression multiplied by $\cos(\phi)$. This is
rather natural as in-phase ALNLS solitons are expected to attract
each other, while out-of-phase ones are expected to experience mutual
repulsion \cite{boris,todd,ricardo}.

\item {\em Retrieving the continuum limit}: In the continuum limit,
one should recover the relevant 
expression of \cite{afan} (see also \cite{avadh}).
In reshaping our expression to match the latter, we observe
that in the continuum limit the discrete spacing $s$ will
appropriately renormalize to the continuum spacing, while 
$\sinh(\beta) \rightarrow \beta$ for the continuum soliton.
Finally, our prefactor of $32$ in Eq.~(\ref{dmeq9}) will match the 
corresponding prefactor
of $8$ in~\cite{avadh,afan}, since the momentum is defined with a factor 
of $1/2$ (as in Eq. (11) of ~\cite{avadh}) and there is an
extra factor of $1/2$ multiplying the RHS of Eq. (\ref{dmeq2})
to match it with the (continuum limit of the) 
RHS of the equations used in~\cite{avadh,afan}.

\end{itemize}

\section{Kink Interactions in a Discrete $\phi^4$ Model}

We  consider now a rather different class of models, namely 
Klein-Gordon lattices. In particular, we focus on a discretization
of the $\phi^4$ field theory of the form:
\begin{eqnarray}
\ddot{u}_n = C \Delta_2 u_n + 2 u_n - u_n^2 (u_{n+1}+u_{n-1}),
\label{dmeq16}
\end{eqnarray}
where $C=1/h^2$ is the coupling parameter and $\Delta_2 u_n=
(u_{n+1}+u_{n-1}-2 u_n)$ stands for the discrete Laplacian.
This type of discretization for the RHS was discussed in~\cite{bender} 
(motivated by ODE examples),
and independently 
rediscovered for the $\phi^4$ field theory 
in~\cite{kevrek,cooper}. An interesting feature
of the model is that, while non-integrable, it possesses
exact solitary wave solutions in the form 
\begin{eqnarray}
u_n= \tanh(\alpha (n - s)) ,
\label{dmeq17}
\end{eqnarray}
where $s$ denotes the center of the kink, $\alpha$ is the inverse width 
of the kink and $\cosh(2 \alpha)=(1+h^2)/(1-h^2)$. This feature, while
useful, is again not an indispensable one for the development
of the Manton procedure. In fact, the only related 
property that is necessary
for the calculation is the exponential tail of the solitary waves
(which can, in fact, be derived by an appropriate exponential
tail ansatz, even if it is not explicitly available in the form
of an analytical solution).

Another property of this model, more crucial for our considerations, 
is that it possesses a discrete analog of the continuum
momentum conservation law. In particular, as  shown in~\cite{kevrek}, 
the relevant momentum is of the form:
\begin{eqnarray}
P=-\frac{1}{2} \sum_{n=-\infty}^{\infty} \dot{u}_n 
\left( u_{n+1} - u_{n-1} \right)\,.
\label{dmeq18}
\end{eqnarray}
One can then perform a similar calculation of $dP/dt$ summing
from $n=M$ to $n=N$, thus obtaining:
\begin{eqnarray}
\frac{dP}{dt}= -\frac{1}{2} \sum_{n=M}^{N} \left[
H(u_{n+1},u_n)-H(u_n,u_{n-1})
+ \dot{u}_n \dot{u}_{n+1} - \dot{u}_{n-1} \dot{u}_n \right] ,
\label{dmeq19}
\end{eqnarray}
where
\begin{eqnarray}
H(u_{n+1},u_n)= C \left(u_{n+1}-u_n\right)^2 + 2 u_n u_{n+1}
- u_n^2 u_{n+1}^2\, . 
\label{dmeq20}
\end{eqnarray}
Given the telescopic nature of the summation, we can easily
infer that 
\begin{equation}
\frac{dP}{dt}=-\frac{1}{2} \left[H(u_{N+1},u_N)-H(u_M,u_{M-1})
+ \dot{u}_N \dot{u}_{N+1} - \dot{u}_{M-1} \dot{u}_M \right]\,.
\label{dmeq21}
\end{equation}

We now use the kink-antikink ansatz of the form (see also
\cite{avadh,manton})
\begin{eqnarray}
u_n=u_n^{(1)} + u_n^{(2)} + r\,,
\label{dmeq22}
\end{eqnarray}
where $r$ is the value of the inhomogeneous background steady
state on which the kinks exist. In the case of the $\phi^4$
model, it can be easily inferred that $r=\pm 1$. Without loss
of generality we use $r=-1$ here. The asymptotic form of the kink
profiles, using the expression of Eq. (\ref{dmeq17}) 
at $n=N$, will be:
\begin{eqnarray}
u_N^{(1)}=1-2 \exp(-2 \alpha N)\,,
\label{dmeq23}
\\
u_N^{(2)}=1-2 \exp\left(2 \alpha (N-s)\right)\,.
\label{dmeq24}
\end{eqnarray}
Notice that in general the first term in the RHS of Eqs. 
(\ref{dmeq23})-(\ref{dmeq24}) is $-r$.

Using the  ansatz 
of Eq. (\ref{dmeq22}) along with Eqs. (\ref{dmeq23})-(\ref{dmeq24})
in the expression of Eq. (\ref{dmeq21}), and isolating the
contributions stemming from the tail-tail interaction 
(to leading order, once again coming solely from $n=N$), we obtain that
\begin{eqnarray}
\frac{dP}{dt} \approx 16 \exp(-2 \alpha s)
\left[ C \sinh^2(\alpha) + \cosh^2(\alpha) \right]\, . 
\label{dmeq25}
\end{eqnarray}

We again make a number of comments on the calculation.

\begin{itemize}

\item {\em Equation of motion for $s(t)$}: In this model, we do not have
an immediately available definition of the soliton mass $M_s$. 
However, considering the momentum itself, we realize that using 
a time dependent center of mass $s(t)$ for the position of the
center of one of the solitons, and substituting that along with
Eq. (\ref{dmeq17}) in Eq. (\ref{dmeq18}), we obtain 
\begin{eqnarray}
\hspace{-20mm}
P= \alpha \dot{s} \sum_{n=-\infty}^{\infty} 
{\rm sech}^2(\alpha (n-s)) 
\left[\tanh\left(\alpha (n+1-s)\right)-\tanh\left(\alpha
    (n-s)\right)\right]\, . 
\label{dmeq26}
\end{eqnarray}
Since $\alpha$ is an effective spacing parameter that will
be used to renormalize the distance in the continuum limit (see
below), we will consider the expression multiplying
$\alpha \dot{s}$ in the above formula of the momentum as the
mass of the kink $M_k$. $M_k$ and $\alpha$ have been computed numerically
for various spacings $h$ in the interval
$(0,1)$ and are shown in Fig.~\ref{dmfig2}. It can be clearly
observed that $M_k$ tends to its continuum limit of $4/3$
($=\int_{-\infty}^{\infty} u_x^2 dx=
\int_{-\infty}^{\infty} {\rm \sech(x)}^4 dx$), validating
the usefulness of this definition. Note also in
the same graph that the behavior of $\alpha$ is 
linear in $h$ for small enough $h$ (in fact, the relevant 
Taylor expansion is $\alpha \approx h + h^3/3 + \dots$ for
small $h$). Let us also mention  that while the expression
of Eq.~(\ref{dmeq17}) for the exact solution appears to be 
necessary for the computation of the kink mass from Eq.~(\ref{dmeq26}),
this constraint can be relaxed as well. For instance, one can quite efficiently
approximate the mass numerically
(see also our numerical calculations below).

\begin{figure}
\begin{center}
\begin{tabular}{c}
    \includegraphics[width=\middlefig]{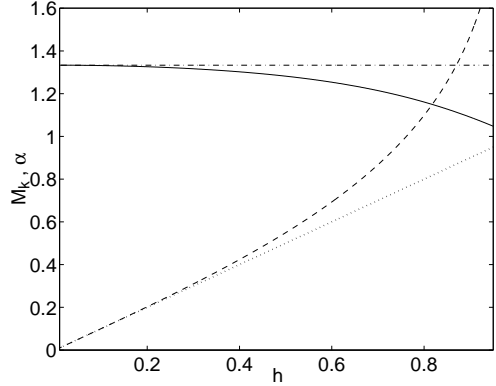} 
\end{tabular}
\caption{The graph shows the dependence of the kink mass $M_k$ 
given by Eq. (\ref{dmeq26}) (solid line)
as a function of the
lattice spacing $h$. The continuum limit corresponding to  $M_k=4/3$ 
(see the main text) is shown by the dash-dotted line. The dashed line
illustrates
the dependence of the parameter $\alpha$ on the spacing $h$, while
the dotted line shows
the linear behavior for comparison.}%
\label{dmfig2}
\end{center}
\end{figure}

Equating the right hand side expression from the momentum of
Eq. (\ref{dmeq25}) [doubled and with a minus sign for reasons 
similar to those mentioned previously for Eq. (\ref{dmeq10})],
with the expression stemming from the
time derivative of $P=\alpha \dot{s} M_k$, we obtain the
evolution dynamics
\begin{eqnarray}
\ddot{s}=-\frac{32 \exp(-2 \alpha s)}{M_k \alpha}
\left[ C \sinh^2(\alpha) + \cosh^2(\alpha) \right].
\label{dmeq27}
\end{eqnarray}
The relevant effective tail-tail kink-antikink interaction 
potential  (for a unit mass particle) can then be obtained in the 
form
\begin{eqnarray}
V(s)=-\frac{16 \exp(-2 \alpha s)}{\alpha^2}
\left[ C \sinh^2(\alpha) + \cosh^2(\alpha) \right]\,.
\label{dmeq28}
\end{eqnarray}
These predictions are tested against numerical simulations, obtaining
once again excellent agreement almost up to the collision point
(the two trajectories only separate visibly for $t \approx 45$,
when the distance of the kink and anti-kink is less than 5 sites). 
The results are illustrated in Fig.~\ref{dmfig3}.

\begin{figure}
\begin{center}
\begin{tabular}{c}
    \includegraphics[width=\middlefig]{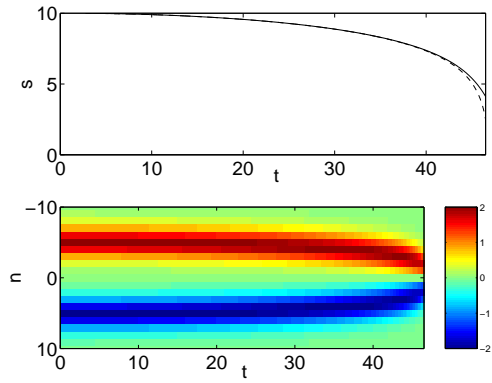} 
\end{tabular}
\caption{Same as Figure 1, but now for the kink-antikink
interaction of the $\phi^4$ model for $h=0.5$. The bottom panel shows
the space-time contour plots of the quantity $D_n=u_{n+1}-u_{n-1}$
whose square we used for our local (approximate) calculation of the
centers of mass of the two kink and antikink, respectively.}%
\label{dmfig3}
\end{center}
\end{figure}

\item {\em Retrieving the continuum limit}: In the continuum limit,
we have to compare Eq.~(\ref{dmeq25})
with the expression for $dP/dt$ that was obtained in~\cite{manton},
\begin{eqnarray}
\frac{dP}{dt} \approx 2 A^2 m^2 \exp(-m s)\,,
\label{dmeq29}
\end{eqnarray}
for two solitons with the  asymptotic  form
\begin{eqnarray}
u^{(1)} &=& -r + A \exp(-m x),
\label{dmeq30}
\\
u^{(2)} &=& -r + A \exp(m (x-s)).
\label{dmeq31}
\end{eqnarray}
Hence, when renormalizing the distance $x$  by our $\alpha$ close to the
continuum limit of $h \rightarrow 0$, and with  $m=2$ and $A=-2$, one obtains
$dP/dt \approx 32 \exp(-2 s)$ for the $\phi^4$ field theory.
We retrieve this result from the expression of Eq.~(\ref{dmeq25}),
since $\lim_{h \rightarrow 0} C (\sinh(\alpha))^2 =1$ and 
$\cosh(2 \alpha) \rightarrow 1$ as $h \rightarrow 0$, by a 
straighforward limit in the expression defining $\alpha$
(just below Eq.~(\ref{dmeq17})). Once again, the correct continuum
limit is obtained from the discrete expression as a special case.
\end{itemize}

\section{Conclusions}

In this work, we have presented a self-contained description
of the way to use the technique first proposed by Manton in \cite{manton}
for the continuum Klein-Gordon models, in order to quantify
tail-tail interactions of soliton solutions of discrete, dispersive nonlinear wave
equations. We demonstrated the calculation via two prototypical
examples, one in the form of soliton-soliton interactions in
the completely integrable Ablowitz-Ladik model and one in the 
form of kink-antikink interactions in a (non-integrable) 
Klein-Gordon chain. We highlighted the key necessary ingredients
for carrying out such a calculation, namely the existence of
a momentum-like quantity which is {\it conserved} due to its 
``telescopic property''. This allows, similarly to the continuum
limit, to arrive at an expression involving only boundary terms
when computing the force exerted on a fraction of the lattice
encompassing one of the two solitary waves. Assessing the
leading order contribution from these surface terms and equating
it (with appropriate prefactors taking into consideration the
total magnitude and direction of the relative change of displacement
of the two waves) to the particle-like momentum of the coherent
structures, we obtain the kinematics of their relative displacement.
The relevant expression clearly highlights the exponential nature
of the tail-tail interactions (which is natural, given the 
exponential tails of the solitary waves). The obtained expressions
have been tested both against the limiting case of their continuum
counterparts which were previously available, and against direct
numerical simulations, providing excellent agreement with the direct
numerical experiments, practically up to the wave collision point
(for the attractive interactions considered herein).

The success of our predictions against the corresponding 
numerical simulations naturally raises the question of 
generalizations, as well as limitations of the
technique. The most significant limitation 
is that ``standard'' discretizations often defy
the existence of a momentum conservation law,
and do not share such a property with their continuum siblings,
restraining themselves to the integer shift invariance, rather
than an effective continuum-like translational invariance. 
It then takes either integrable (such as e.g. the ones
of \cite{AL}) or ``non-standard'' (see e.g. \cite{kevrek,cooper}
and references therein) discretizations to circumvent this
difficulty and possess a discrete analog of the momentum.
Hence, it would be very desirable to be able to extend
our considerations to cases where the momentum conservation
is absent, hopefully obtaining, in addition to the tail-tail
terms already captured above, the ones from the local, periodic 
Peierls-Nabarro barrier. Another natural extension of the Manton 
approach (in fact, both in the continuum and in the discrete systems) 
would be the study of its multi-dimensional analog. While by no
means straightforward (especially given the sparsity of
analytically available solutions and the difficulty of
questions such as the definition of the right momentum/contour),
it constitutes an increasingly relevant generalization of the
considerations presented herein.
Work along these directions is currently underway and will
be reported in future publications.

\vspace{5mm}

This work was supported in part by the U.S. Department of Energy. 
PGK is grateful to the Eppley Foundation
for Research, the NSF-DMS-0204585, NSF-DMS-0505063 
and the NSF-CAREER program for 
financial support and to the Center for Nonlinear Studies
of Los Alamos National Laboratory for its hospitality.
I.B. acknowledges support of the Swiss National Science Foundation 
and  the Center for Nonlinear Studies
of Los Alamos National Laboratory for its hospitality.

\end{document}